\documentstyle{article}

\newcommand{\beq}{\begin{equation}}
\newcommand{\eeq}{\end{equation}}
\begin{document}

\setlength{\topmargin}{-0.8cm}
\addtolength {\oddsidemargin} {-2.2cm}
\setlength{\parskip}{1pc}
\setlength{\parindent}{0pc}

\begin{center}
{\Large \bf Deterministic SR in a Piecewise Linear Chaotic Map}

\vspace{0.5cm}
{\bf Sitabhra Sinha ${}^1$ and Bikas K. Chakrabarti ${}^2$}

${}^1$ Machine Intelligence Unit, Indian Statistical Institute,
Calcutta 700 035, India.\\
${}^2$ Saha Institute of Nuclear Physics,
1/AF Bidhan Nagar, Calcutta - 700 064,
India.
\end{center}

\begin{abstract}
The phenomenon of Stochastic Resonance (SR) is observed in
a completely deterministic setting - with thermal noise 
being replaced by one-dimensional chaos.
The piecewise linear map investigated in the paper
shows a transition from symmetry-broken to symmetric chaos on increasing
a system parameter. In the latter state, the chaotic
trajectory switches between the two formerly disjoint attractors,
driven by the map's inherent dynamics. This chaotic switching rate is 
found to `resonate' with the frequency of an externally applied
periodic perturbation (multiplicative or additive). By periodically
modulating the parameter at a specific frequency $\omega$ we observe the
existence of resonance where the response of the system (in terms of the
residence-time distribution) is maximum. This is a clear indication
of SR-like behavior in a chaotic system.\\

\end{abstract}

``Stochastic Resonance'' (SR) is a recently observed nonlinear phenomena 
in noisy
systems, where the noise helps in amplifying a sub threshold signal (which
would have been otherwise undetected) when the signal frequency is close to
a critical value \cite{Benzi81}. This occurs because of noise-induced
hopping between multiple stable states of a system, locking on to an
externally imposed periodic signal. A theoretical understanding of SR in 
bistable systems, subject to both periodic and random forcing,
has been obtained based on the rate equation
approach \cite{McNamara}. As the output of a
chaotic process is indistinguishable
from that of a noisy system, the question of whether a similar process
occurs in the former case has long been debated. In fact, Benzi {\it et al}
\cite{Benzi81} indicated that the Lorenz system of equations, a well-known
paradigm of chaotic behavior might be
showing SR. Later studies \cite{Nicolis},
\cite{Anish} in both discrete-time and continuous-time systems seemed to
support this view. However, it is difficult to guarantee that the response
behavior is due to ``resonance'' and not due to ``forcing''.
In the latter case, the periodic perturbation is of so large an
amplitude, that the system is forced to follow the driving frequency
of the periodic forcing. The ambiguity is partly because the
Signal-to-Noise Ratio (SNR) is a monotonically decreasing function of
the forcing
frequency and cannot be used to distinguish between resonance and forcing.
In the present work this problem is avoided by measuring the response of
the system in terms of the normalized distribution of residence times 
\cite{Gamm95}. For SR, this measure shows non-monotonicity with the
variation of both noise intensity and signal frequency. 

Ippen {\it et al} \cite{Ippen} have used a chaotic driving term to
show SR-like behavior in the SNR of the system response.
However in this case
the chaos is supplied from outside, and not inherent to the system.
If SR is indeed used for information processing by biological systems,
then it is likely that organs producing chaotic behavior might enhance
their survival capability through selective amplification of signals
in a noisy background. In this case, the inherent chaos of the system
itself could play the role of ``noise''. In the model proposed in this paper,
a simple one-dimensional map has been shown to use its inherent chaoticity
to replicate SR-like phenomena. This suggests a deep relation
between stochastic resonance on the one hand, and crises in chaotic dynamics
on the other, mentioned in \cite{Carroll}. The present work also supports
this view.

The simplest chaotic system to show SR-type behavior are one-dimensional
maps with two critical points. The most commonly studied system of this kind
is the cubic map \cite{cubic},
$$
x_{n+1} = a x_{n}^3 + (1 - a) x_{n},
$$
where $a$ is a tunable parameter. The map is found to consist of two
attractors, the initial condition determining the attractor into which
the system settles. Various properties
of such `bimodal' maps differ from those
observed for the well-studied class of maps with a single critical point
(e.g., the logistic map). 

Recently, SR has been studied in 1-D maps with two well-defined states 
(but not necessarily stable) with switching between them aided
by either additive or multiplicative external noise \cite{Gade}.
However, dynamical contact of two chaotic 1-D maps can also induce
rhythmic hopping between the two domains of the system \cite{Seko}.
The present work shows how the
chaotic dynamics of a system can itself be used for resonant switching
between two states, without introducing any external noise.

The model chosen here is a piecewise linear bimodal map,
henceforth referred to as the Discontinuous
Anti-symmetric Tent (DAT) map, defined in the interval [-1,1]:
\begin{equation}
x (n+1) = {\rm F}(x_n) \left\{
\begin{array}{ll}
  1 + a(0.5-x(n)),  & {\rm if}~~x(n) \geq 0.5\\
  1 - a(0.5-x(n)),  & {\rm if}~~0 < x(n) < 0.5\\
 -1 + a(0.5+x(n)), & {\rm if}~~-0.5 < x(n) < 0\\
 -1 - a(0.5+x(n)),  & {\rm if}~~x(n) \leq -0.5.
\end{array}
\right.
\end{equation}
The map has a discontinuity at $x = 0$. The behavior of the system was
controlled by the parameter $a$, $(0<a<4)$. Onset of
chaos occurs at $a=1$. The chaos is symmetry-broken,
i.e., the trajectory is restricted
to either of the two sub-intervals R:(0,1] and L:(0,-1], depending on
initial condition. Symmetry is restored at $a=2$. The lyapunov exponent
of the map is a simple monotonic function of the parameter $a$. The
piecewise-linear nature of the map makes its behavior simpler to study
than, say, the cubic map described above.
The map is shown in fig. 1, the inset giving a detailed
picture of the region around the discontinuity at $x = 0$.
Fig. 2 shows the evolution
of the map's attractor with $a$ increasing from 0 to 4.

The map has a symmetrical pair of fixed points $x^*_{1,2}= \pm {\frac
{1+a/2}{1+a}}$
which are stable for $0<a<1$ and unstable for $a>1$. Another pair of
unstable fixed points, $x^*_{3,4} = \pm {\frac{1-a/2}{1-a}}$ come into
existence for $a>2$. It is to be noted that as $a \rightarrow 2$ from above,
$x^*_{3,4}$ both collide at $x=0$ causing an interior crisis, which
leads to symmetry-breaking of the chaotic attractor.

To observe SR, the value
of $a$ was kept close to 2, and then modulated sinusoidally with amplitude
$\delta$ and frequency $\omega$, i.e.,
\begin{equation}
a_{n+1} = \left \{
\begin{array}{ll}
a_{0} ~+~ \delta~ {\rm sin}(2 \pi \omega n), & {\rm if}~~x \in {\rm R}\\
a_{0} ~-~ \delta~ {\rm sin}(2 \pi \omega n), & {\rm if}~~x \in {\rm L}.
\end{array}
\right.
\end{equation}
We refer to this henceforth as multiplicative or parametric perturbation, to
distinguish it from additive perturbation (discussed later).

The system immediately offers an analogy to the classical bistable well
scenario of SR. The sub intervals L and R correspond to the two wells between
which the system hops to and fro, aided by the inherent noise (chaos) and
the external periodic forcing. In each positive (negative) half-cycle of the
periodic signal, a portion of the map defined over R (L) overlaps
into the domain of the other portion defined over L (R). This is analogous
to the successive raising and lowering of the wells in synchronization
with the signal frequency, allowing the system to escape
from one well to the other. The resultant intermittent switching of
the trajectory between L and R is shown in Fig. 3.
If the dynamics of the system due to the internal noise (chaos) has some
inherent time-scale (say $n_k$),
as ${\frac{1}{\omega}} \rightarrow n_k$
the two time-scales may lock onto each other. This resonance should be
observable through an increase in the response characteristics of the map.

The response of the system is measured
in terms of the normalized distribution
of residence times, $N(n)$ \cite{Gamm95}. This distribution shows a series
of peaks centered at $n_j = (j - {\frac{1}{2}}) n_0$, i.e.,odd-integral
multiples of the forcing period, $n_{0} = {\frac{1}{\omega}}$. 
The strength of the $j$-th peak
\begin{equation}
P_j = \int_{n_j - \alpha n_0}^{n_j
+ \alpha n_0} N(n) dn ~~~(0<\alpha<0.25),
\end{equation}
is obtained  at different values of $\omega$, keeping $a_0$
fixed for $j$=1,2 and 3. To maximize sensitivity, $\alpha$
was taken to be $0.25$.
For $a_0=2.01$ and $\delta=0.05$, the response of the system showed a
non-monotonic behavior as $\omega$ was varied, with $P_1$ peaking
at $\omega_1 \sim 1/400$, a value dependent upon $a_0$ $-$ a clear
signature of SR-type phenomenon. $P_2$ and $P_3$ also showed non-monotonic
behavior, peaking roughly at odd-integral multiples of $\omega_1$
(Fig. 4(a)).

Similar observations of $P_j$ were done also by varying $a_0$
keeping $\omega$ fixed. Fig. 4(b) shows the results of simulations
for $\omega = 1/400$ and
$\delta=0.05$. Here also a non-monotonicity was observed
for $P_1$,$P_2$ and $P_3$.
The broadness of the response curve and the magnitude of the
peak-strengths
are a function of the perturbation magnitude, $\delta$. The variation of
$P_1$ with $a_0$ for different values of $\delta$ were also studied.
As $\delta$ decreases, the response curve becomes more
sharply peaked while the peak-strength decreases. 

Note that, the parametric perturbation cannot be done without
modulating the noise-intensity. This seems to be the
principal difference between this type of `chaotic resonance' and
classical SR seems to be  As the local slope of the map, $a$, is varied
periodically, the internal noise, whose intensity is a function of the
lyapunov exponent (and hence of $a$) also varies periodically. In contrast,
for classical SR, the wells are raised or lowered periodically without
affecting the external noise, which is independent of the geometry
of the wells.

Analytical calculations were done to obtain the invariant probability density
and the dominant time-scale governing the residence-time distribution. This
was done by proper partitioning of the domain of definition of the system
and obtaining the eigenvalues of the corresponding transition matrix. 
From Fig. 2, it is clear that the system spends a longer time in the interval
[$-\epsilon/2, \epsilon/2$], where $\epsilon=a_0 - 2$. So a natural
partitioning of the interval [-1,1] is into the four sub-intervals:
$C_1: [- 1,- \epsilon/2]$, $C_2: [- \epsilon/2, 0]$, $C_3: [0, \epsilon/2]$
and $C_4: [\epsilon/2, 1]$. This is an exactly Markov partition at
integral values of $\epsilon$, i.e., the partition boundaries, $\{p_i\}$ 
transform into each other on application of the map dynamics,
( $f(p_j) \in \{ p_i \}$). It is assumed that for $a_0 \rightarrow 0$
the partitioning approximately retains its Markovian character, so that
the process can be mapped onto a Markov process. Close to $\epsilon = 0$,
the transition matrix corresponding to the above partitioning is:
\begin{equation}
W = \begin{array} {|cccc|}
{\frac{1-{\epsilon}/2-{\epsilon}^2/4}{1-{\epsilon}^2/
4}} & {\frac{\epsilon}{4(1-{\epsilon}^2/4)}} & {\frac{\epsilon}{
4(1-{\epsilon}^2/4)}} & 0\\
{\frac{\epsilon}{2+\epsilon}} & {\frac{1}{2+\epsilon}} &
{\frac{1}{2+\epsilon}} & 0\\
0 & {\frac{1}{2+\epsilon}} & {\frac{1}{2+\epsilon}} & 
{\frac{\epsilon}{2+\epsilon}}\\
0 & {\frac{\epsilon}{4(1-{\epsilon}^2/4)}} & 
{\frac{\epsilon}{4(1-{\epsilon}^2/4)}} &
{\frac{1-{\epsilon}/2-{\epsilon}^2/4}{1-{\epsilon}^2/4}}
\end{array}
\end{equation}
where, $W_{ij} = P(C_i, C_j)$ is the probability of transition
from $C_i$ to $C_j$.
The eigenvalues of the above matrix are $\lambda_1 = 1$,
$\lambda_2 = {\frac{1-{\epsilon}/2-{\epsilon}^2/4}{1-{\epsilon}^2/4}}$,
$\lambda_3 = {\frac{1-{\epsilon}}{1-{\epsilon}^2/4}}$ and $\lambda_4 =0$.
The largest
eigenvalue, 1, corresponds to the invariant probability density over
the four intervals. The next largest
eigenvalue dominates any time-dependent phenomena. The relevant time-scale
(i.e., the mean residence time) is given by
\begin{equation}
n_k = {\frac {-1}{{\log}({\frac{1-{\epsilon}/2
-{\epsilon}^2/4}{1-{\epsilon}^2/4}}) }} \simeq
{\frac{-1}{\log(1-\epsilon/2)}}.
\end{equation}
So, for $a_0=2.01$, $n_k \simeq 200$. 
This predicts that a peak in the response should be observed at a frequency
${\frac{1}{2 n_k}} \simeq 1/400$, which agrees with the simulation results.
For small $\epsilon$, $\lambda_2 \simeq {\rm exp}(-\epsilon /2)$.
Therefore, as $a_{0} \rightarrow 2$ from above, 
the residence time diverges as 
\beq
n_k \sim (a_{0} - a_{0}^{*})^{- 1}, ~~~a_{0}^{*} = 2.
\eeq

The mean time spent by the trajectory in any one of the 
sub-intervals (L or R) can be
calculated exactly for piecewise linear maps \cite{Everson}. 
For $\epsilon$ > 0, the intervals $\beta_1$ = ($0, {\frac{\epsilon}{2
(2 + \epsilon)}}$] and $\beta_2$ = [$1-{\frac{\epsilon}{2
(2 + \epsilon}}, 1$] of R maps to L, so that
the trajectory escapes from one sub-interval to the other. Note the
symmetrical placement of the two R $\rightarrow$ L `escape regions' about
$x=0.5$, because of the symmetry ${\rm F}(1/2 - x) = {\rm F}(1/2 + x)$
of the DAT map. So the total fraction of R escaping to L after one 
iteration is $l_1= {\frac{2 \epsilon}{2 (2 + \epsilon)}}$.
Let us now consider the first pre-image of $\beta_1$ and $\beta_2$, which
escapes from R to L after two iterations. The total fraction of R belonging
to this set is $l_2= {\frac{4 \epsilon}{2 (2 + \epsilon)^2}}$. Proceeding
in this manner, we find from the geometry of the map that the total
fraction of R which maps to L after $n$ iterations is
\beq
l_n = {\frac{2^n \epsilon}{2 (2 + \epsilon)^n}}.
\eeq
These are just the probabilities that the trajectory spends a period
of $n$ iterations in R before escaping to L ($\sum_{j=1}^{\infty}
l_j = 1$). So the average lifetime of a trajectory
in R is
\beq
<n> = \sum_{j=1}^{\infty} (j-1) l_j = {\frac{2}{\epsilon}}.
\eeq
For $a_0$ = 2.01, $< n >$ = 200, in good agreement with the result
obtained using the approximate Markov partitioning. The above equation
also establishes exactly the linear scaling relation of the mean lifetime
about $\epsilon = 0$, with $< n >$ diverging at $a_0 = 2$. By symmetry
of the map, identical results will be obtained if we consider
the trajectory switching from L to R.

Another interesting quantity which also shows a scaling behavior
around $\epsilon$ = 0, is the drift rate, $v$, from one sub-interval
to the other \cite{Grossman}. This measures the rate at which
the chaotic trajectory 
switches between L and R. Owing to the symmetry ${\rm F}(-
x) = - {\rm F}(x)$ of the
DAT map, the net drift rate is zero, i.e., switching to either
sub-interval occurs equally often. Let us consider switching from R to L
(identical results will hold for switching in the opposite direction
due to symmetry). The drift rate is measured by the fraction of R mapping
to L per iteration. Hence,
\beq
v = {\frac {\epsilon}{2 + \epsilon}}.
\eeq
It is again a linear scaling relation as $a_0 \rightarrow 2$ from above.
Note that, $a_0 < 2$, $v = 0$ as the two sub-intervals are isolated from
each other. 
Thus $v$ is analogous to an `order parameter', having a
finite (positive) value above $a_0=2$ and zero below it. 
This suggests that the merging of the chaotic attractors at $a_0 =2$
is akin to a critical phenomena, with the local slope $a_0$ as the
tuning parameter.

Similar study was also conducted with additive perturbation for
the above map. In this case the dynamical system is defined as follows:
\beq
x_{n+1}={\bf \rm F}(x_{n}) ~+~ \delta {\rm sin}(2 \pi \omega n).
\eeq
For $a = 1.9$ (say), the map has
two disconnected sub-intervals, L:[-1,0) and R:(0,1]. However, an additive
perturbation of magnitude $\delta > 0.1$ causes a portion of L
to diffuse into R in the positive half-cycle of the sinusoidal signal
(of frequency $\omega$). Similarly, in the negative half-cycle, a portion of
the R interval diffuses into L. The long-term behavior of the map is
described by a ``smeared-out'' DAT map with a
width $\delta$ rather than the ``crisp'' 
piecewise linear DAT map with $a_{0}=1.9$. This happens as the map performs 
a periodic vertical motion, causing a smearing-out over time. 
The simulation results showed non-monotonic behavior for the response as
either $\omega$ or $a_{0}$ was varied, keeping the other constant, but
this was less marked than in the case of multiplicative perturbation.
This work can be seen in context with studies conducted on the dynamics
of the logistic map under parametric perturbation \cite{Nand96}.

Low-dimensional discrete-time dynamical
systems are amenable to several analytical
techniques and hence can be well-understood compared to other systems.
The examination of resonance phenomena in this scenario was for ease of
numerical and theoretical analysis. However, it is reasonable to assume that
similar behavior occurs in higher-dimensional chaotic system, described by
both maps and differential equations. 

The close resemblance of the merging of attractors with critical phenomena
has possible relevance to SR in Ising systems. Although numerical studies
have reported SR in kinetic Ising system, it seems to be inconclusive
as the primary peak strength of the
normalized residence-time distribution shows
only a monotonic behavior \cite{Sides}. This response profile is identical
to that observed in DAT Map for $a_0 < 2$. A study of kinetic aspects like
hysteresis is planned to be undertaken, which should give information
concerning the phase-dependence of the resonance behavior.

The observation of `SR' in chaotic systems also has implications for the
area of noisy information processing. It has been proposed that the
sensory apparatus of several creatures use SR to enhance their sensitivity
to weak external stimulus, e.g., the approach of a predator.
Some experimental
work on crayfish have provided supporting evidence to this assertion
\cite{Douglass}.
The above study indicates that external noise is not necessary for such
amplification as chaos in neural networks can enhance weak signals.
As chaotic behavior is extremely common in a recurrent network
of excitatory and inhibitory neurons, such a scenario is not entirely
unlikely to have occurred
in the biological world. This can however be confirmed only by further
biological studies and detailed modeling of the phenomena. 

Several interesting comments
on the work were made by Prashant M. Gade (JNCASR, Bangalore).
Jayanta K. Bhattacharjee (IACS, Calcutta) made some useful suggestions.

\vspace{1cm}
\begin{center}
\Large{Figure Captions}
\end{center}

{\bf Fig. 1} The DAT map for $a_0 = 2.01$. Inset: a magnified view of the
map in the interval $[-0.005, 0.005] \times [-0.005, 0.005]$.

{\bf Fig. 2} Attractor of the DAT map versus $a_0$. The figure was
obtained for $x_0 \in {\rm R}$. For $x_0 \in {\rm L}$ the corresponding
image is obtained by reflecting about $x$-axis.

{\bf Fig. 3} The time-evolution of the sinusoidally perturbed DAT map
for $a_0 = 2.01$, $\omega = 1/400$ and $\delta = 0.05$. The broken line is
the boundary between L and R.

{\bf Fig. 4} (a) $P_n$ ($n$ = 1, 2, 3) versus $\omega$ for $a_0 = 2.01$ and
$\delta = 0.05$, (b) $P_n$ ($n$ = 1, 2, 3) versus $a_0$ for $\omega =
1/400$ and $\delta = 0.05$. The circles represent the average value of
$P_n$ for 18 different initial values of $x$, the bars representing the
standard deviation. The data points are joined by solid lines for the
reader's convenience.

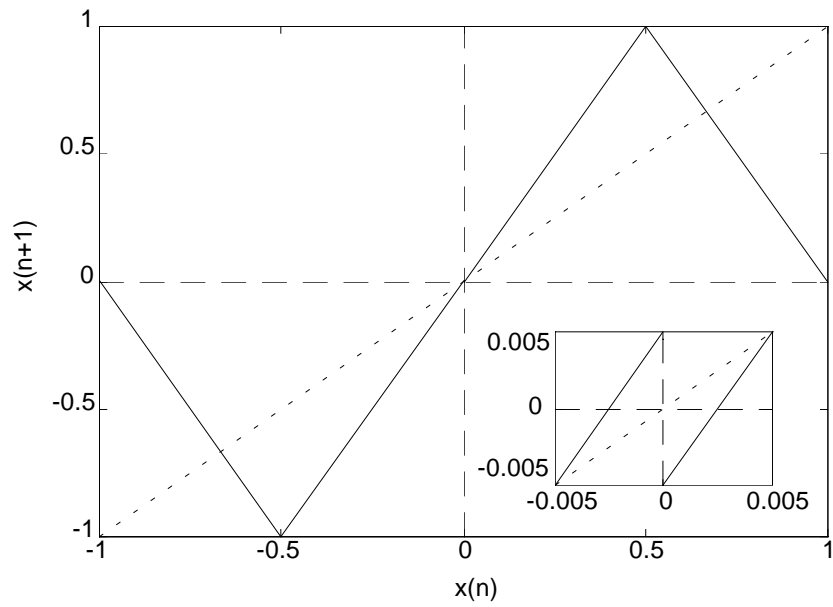
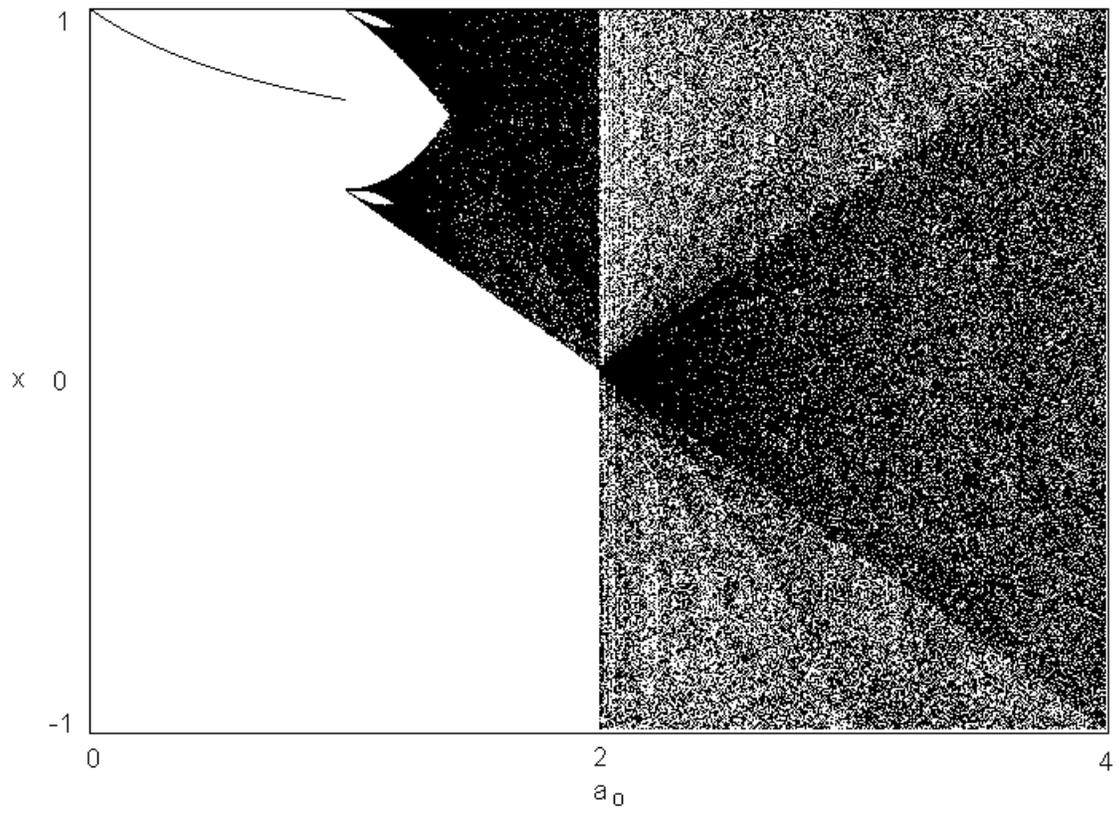
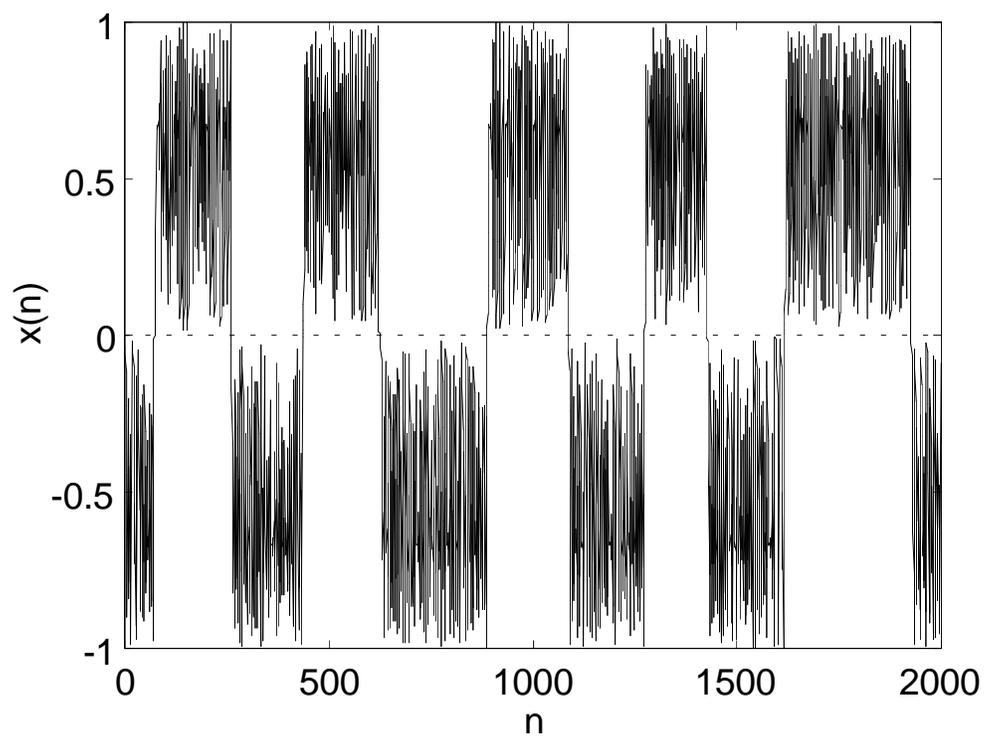
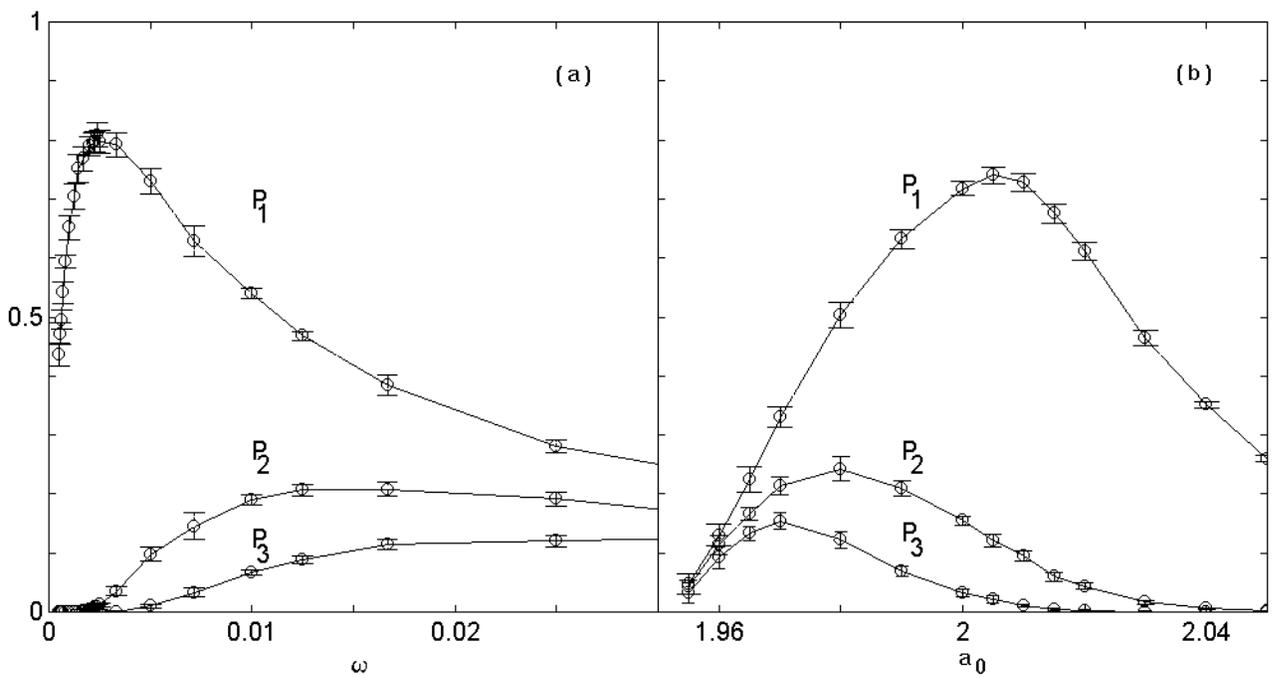
\end{document}